# Phosphorus cluster cations formed in doped helium nanodroplets are different

Simon Albertini[1,2b +], Faro Hechenberger[1+], Siegfried Kollotzek[1], Lukas Tiefenthaler[1], Paul Martini[1], Martin Kuhn[1], Alexander Menzel[3], Elias Jabbour Al Maalouf[4], Harald Schöbel[2], Masoomeh Mahmoodi-Darian[5*], and Paul Scheier[1]

[1] University of Innsbruck, Institute for Ion Physics and Applied Physics, Austria
[2] Management Center Innsbruck, Department Biotechnology & Food Engineering, Austria
[3] University of Innsbruck, Institute for Physical Chemistry, Austria
[4] Faculté des Sciences IV, Laboratoire Energétique et Réactivité à L'Echelle Nanométrique (EREN), Université Libanaise, Haouch El-Omara, Zahlé, Lebanon
[5] Department of Physics, Karaj Branch, Islamic Azad University, Karaj, Iran
[+] These authors contributed equally
[*] Correspondence should be directed to masoomeh2001@yahoo.co.uk

## Abstract

Positively charged cluster ions of phosphorus were formed upon electron ionization of doped helium nanodroplets. The vapors of red phosphorus and a phosphate sample were picked up into neutral and charged helium nanodroplets. Independent on the conditions used, the cluster size distributions exhibit pronounced odd-even oscillations that are opposite to almost all experimental and theoretical patterns published in the literature. The low temperature environment of the superfluid He matrix quenches fragmentation and the charged phosphorus clusters resemble the structure of the neutral precursors.

## Introduction

Phosphorus has several allotropes that exhibit strikingly different properties [1]. White phosphorus was the first form of the element to be prepared in 1669 [2], and it is still the best characterized. At room temperature, its crystal structure is a loosely bound aggregation of tetrahedral $P_4$ units. Black phosphorus is the thermodynamically stable allotrope at ambient conditions and forms puckered two-dimensional sheets, similar to graphite. The third and most complex allotrope form of phosphorus is red or violet phosphorus. Vaporizing red phosphorus at low temperatures produces a gas primarily consisting of $P_4$. With increasing temperatures, the relative amount of $P_2$ vapor increases and atomic phosphorus vapor begins to appear at even higher temperatures [3]. Due to its low thermodynamic stability the abundance of $P_3$ in the vapor relative to those of $P_2$ and $P_4$ is negligible. Smets et al. [4] studied photo ionization of $P_4$ vapor and determined $P_4^+$ as the dominant product at all photon energies investigated. At a photon energy of 19 eV the ratios between the ion yields of the fragments $P_2^+$ and $P_3^+$ and the parent cation $P_4^+$ are 0.22 and 0.16, respectively. No atomic fragment $P^+$ was observed. Monnom et al. [5] determined partial electron ionization cross sections of $P_4$ and obtained at an electron energy of 70 eV relative abundances for the fragments $P^+$, $P_2^+$ and $P_3^+$ with respect to the parent cation $P_4^+$ of 0.097, 0.2 and 0.074, respectively.

Phosphorus clusters are likely to display equally rich structural variety. This has motivated many experimental and theoretical studies starting with the pioneering mass spectrometric experiments of Martin in 1986 [6]. They vaporized red phosphorus at 300°C into a reaction chamber filled with 1 mbar helium. The outer mantle of the reaction chamber was cooled with liquid nitrogen and neutral clusters were formed via gas aggregation. Positively charged clusters $P_n^+$ were formed upon electron ionization with 80 eV electrons. The recorded mass spectra showed pronounced odd-even oscillations with odd numbered clusters about twice as intense as even numbered species.

Huang et al. [7-9] formed cationic and anionic phosphorus clusters by laser vaporization of red phosphorus without subsequent ionization. Cluster sizes were reported up to $P_{89}^-$ [10] and $P_{49}^+$ [9]. Particularly intense peaks were observed for n = 8k + 1 (k ≥ 5 and k ≥ 3, for anions and cations, respectively). Bulgakov et al. studied phosphorus clusters up to $P_{91}^+$ utilizing mass spectrometry [11-13]. Electron ionization of neutral phosphorus clusters formed upon laser ablation were preferentially even-numbered with local abundance maxima at n = 8, 10, 14, and 40. In contrast, odd-numbered clusters were more abundant for charged clusters formed during the laser ablation process with local abundance maxima at n = 8k + 1 (k ≥ 5). Kong reported on large phosphorus cluster cations up to n = 300 utilizing a MALDI TOF instrument and anions up to n=500 utilizing a Fourier transform ion cyclotron resonance mass spectrometer system [14]. $P_n^+$ cations with n=8k+1 (k=3-10) exhibit again higher intensities. However, this 8k+1 rule became insignificant for k>11. Similar results can be observed for the anions and the 8k+1 rule was applicable for k<18.

The structure of phosphorus clusters in different charge states has been investigated by various quantum chemical methods [15-26] ranging from a molecular dynamics/density functional MD/DF approach [15] to coupled-cluster theory CCSD(T) [19] and comprehensive genetic algorithm DFT [26]. Compared to extensive computational studies of neutral phosphorus clusters [16-19, 21, 22, 26], much less is known about charged clusters, which were most frequently investigated experimentally. For neutral phosphorus clusters, odd-membered clusters $P_{2n+1}$ were calculated to be less stable than the most stable even-membered cluster $P_{2n}$ [16]. For larger cationic phosphorus clusters with n=25+8k (k= 1 - 8), Chen et al. [20] constructed a series of chain-like configurations by adding cuneane $P_8$ units to a presumed $C_s$ structure of $P_{25}^+$. Xue et al. [25] studied the global minimum structures of odd-sized $P_{2m+1}^+$ (m = 1–12) cations using first-principles simulated annealing.

The stability and fragmentation of phosphorus clusters were also experimentally studied via collision-induced dissociation (CID) mass spectrometry [9, 27]. The primary dissociation pathway of $P_{2m+1}^+$ (6 ≤ m ≤ 11) is the loss of a $P_4$ unit. For magic cluster ions of $P_{8k+1}^+$ (3 ≤ k ≤ 8), the dissociation pathways progressively changed from the loss of $P_4$ to loss of $P_8$.

Here we investigate $P_n^+$ cluster ions formed upon pickup into neutral and charged helium nanodroplets (HNDs). The ionization mechanism of dopants in helium nanodroplets depends on their location [28]. Heliophobic dopants, such as alkali atoms [29, 30] or small alkali clusters [31-34] reside in dimples on the surface of the droplets and are most efficiently ionized via Penning ionization by heliophobic metastable helium atoms He*. Heliophilic dopants, such as phosphorus and $P_n$ clusters submerge into the droplets and are preferentially ionized via charge transfer from an initially formed $He^+$ or a small charged $He_m^+$ cluster ion. This process is highly exothermic as the difference in the ionization energy of He ($He_m$) and $P_n$ is completely transferred into the dopant cation. The vertical ionization energy of $P_4$ is 9.5 eV [5, 35, 36] and its adiabatic ionization energy is 9.2 eV [36]. More than 15 eV excess energy remains in the $P_4^+$ ion upon charge transfer from $He^+$ as the ionization energy of He is 24.59 eV [37]. If this energy can be transferred to the surrounding helium matrix before fragmentation occurs, the resulting mass spectra match the neutral dopant distribution. The effect of the low temperature matrix on the resulting cluster size distributions is investigated via high-resolution mass spectrometry.

## Experimental

### XPS measurements

Resbond ® 920 powder was mixed with pure water (14:1). A thin layer of the resulting paste was applied to a stainless steel surface and heated to 150 °C for eight hours in a compartment dryer. This prepared sample was analyzed by XPS using a Thermo MultiLab 2000 spectrometer with an alpha 110 hemispherical analyzer (Thermo Electron) in the constant analyzer energy mode (surveys with pass energy 100 eV, energy resolution 2 eV; detailed spectra with pass energy 25eV, energy resolution 0.8 eV). A twin crystal monochromator provided Al Kα radiation (1486.6 eV) with a focus of 650 µm in diameter.

### Production of helium nanodroplets

Phosphorus clusters were produced in helium nanodroplets using two different experimental setups, called ClusTOF and Toffy. In both machines, HNDs were produced by supersonic jet expansion of ultrapure helium (99.9999 %) through a 5 µm pinhole nozzle into ultrahigh vacuum. The nozzle was cooled with a closed-cycle cryo cooler (Sumitomo Heavy industries) and counterheated with an ohmic resistor operated with a PID controller (Lakeshore Model 331). Without pressurized helium (2 to 2.4 MPa) in the source the pressure behind the nozzle was below $10^{-6}$ Pa and rises during operation with helium to $1 \times 10^{-2}$ Pa. According to Gomez et al. [38] the resulting average droplet size is between $2.5 \times 10^5$ and $1 \times 10^6$ He atoms for the present conditions. To prevent destruction of HNDs by collisions with shock fronts, the resulting jet of He was then passing a molecular beam skimmer (Beam Dynamics, Inc), located downstream of the nozzle.

### ClusTOF

In ClusTOF, the nozzle temperatrue for HND production was 9.99 K with a stagnant He pressure of 2.4 MPa. The beam is then passing a 0.8 mm skimmer positioned about 5 mm from the nozzle. The He beam was then passing a differentially pumped pickup chamber, containig an oven composed from an alumina ceramics bound with RESBOND® 920 Alumina Ceramic Adhesive (Final Advanced Materials GmbH). The ClusTOF mass spectrum shown in this work was recorded during a test run of a new high-temperature, ceramic based oven without any additional sample. When heated to 720 K, the oven started to emit P vapor. XPS measurements showed that the adhesive contains phosohorus compounds that are presumably the source of this vapor. Atomic or molecular phosphorus was picked up by the HNDs and agglomeration lead to the formation of neutral phosphorus clusters inside the helium droplets. Subsequently, the doped HND beam passed a 2.5 mm aperture into the next differentially pumped chamber with a pressure of pressure $3.8 \times 10^{-6}$ Pa containing a Nier-type electron impact ionization unit. This ion source was operated at an electron energy of 80 eV and an electron current of 50 µA. The probability for an electron to hit the dopant cluster is negligibly small compared to the ionization of a He atom. Ion induced dipole interaction attracts the charge to the dopant and charge transfer from $He_x^+$ to $P_n$ ionizes the phosphorus cluster. Low-mass ions are ejected from the large droplets due to mutual Coulomb repulsion of charge centers in multiply-charged HNDs [39]. These ions are extracted by weak electrostatic fields and guided into the extraction region of a reflectron time of flight mass spectrometer (H-Tof, Tofwerke) via a stack of einzel lenses. For further details of this experimental setup see ref. [40].

**Toffy**

In Toffy, HNDs were produced at a nozzle temperature of 9.2 K and a stagnant He pressure of 2.0 MPa. The beam is then passing a 0.5 mm skimmer positioned about 10 mm from the nozzle. In contrast to ClusTOF, the HNDs are ionized by a Nier-type electron impact ionization unit directly after the skimmer. The ion source was operated at an electron energy of 70 eV and an electron current of 200 μA. Charged droplets were then mass per charge filtered for approximately $2.5\times10^5$ He/z before pickup of a vapor formed upon heating red phosphorus (99.99 %, Sigma-Aldrich product number 343242) to 690 K in an ohmically heated oven. The polarizability of capured phosphorus vapor is much higher than of He and therefore, ion induced dipole interaction attracts dopants to the charge centers. Charge transfer from $He_n^+$ to the first dopant arriving is highly exothermic and might lead to fragmentation of molecular dopants, unless the surrounding helium matrix is able to quench the excess energy before the fragments are separating. Further dopants will simply attach to an already existing $P_n^+$ cluster and the binding energy will be transferred to the surrounding He and lead to the evaporation of He atoms from the surface of the HND. In a very recent study, we observed that pre-doping of charged HNDs with a small amount of molecular hydrogen leads to the formation of odd-numberd hydrogen cluster ions $H_{2n+1}^+$ that efficiently transfer a proton to subsequently captured dopants having a higher proton affinity [41]. This is the case for all dopants other than Ne, Ar and $O_2$. Compared to electron transfer, proton transfer is a slow process which enables the surrounding He matrix to quench excess energy, even if it amounts several eV. Thus, proton transfer ionization inside a HND turns out to be a very gentle ionization mechanism that reduces fragmentation completely. After cluster formation, any remaining He atoms were boiled off in an evaporation cell with a He pressure of 47 mPa and the resulting ionic phosphorus clusters were analyzed in a time of flight mass spectrometer (Q-TOF Ultima Waters/Micromass). For further details of this experimental setup see [42].

## Results

**XPS Analysis of Resbond ® 920**

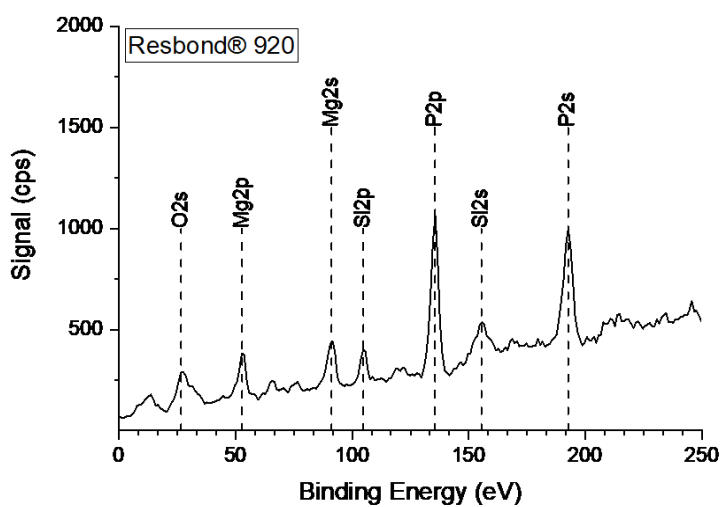

Figure 1: XPS analysis of Resbond® 920 paste applied to a stainless steel surface. Characteristic peaks for O, Mg, Si, and P are labelled according to the calibration with C1s (adventitious carbon) [43].

The XPS analysis of Resbond® 920 as showed in Figure 1 displayed several minor peaks resulting from impurities, such as magnesium and silicon. The spectrum was calibrated using the C1s (adventitious carbon) peak at 284.5 eV [43]. Relativ abundance of O:P (58%:16% ) and the P2p binding energy indicate that the surface of Resbond® 920 mostly contains $P_2O_5$, $P_4O_{10}$ or similar. Reference values for P2p shown are taken from in P2p in $P_4O_{10}$.

## Mass Spectra of phosphorus

Figure 2 shows the spectrum of cluster formation upon doping P from Resbond® 920 Alumina Ceramic Adhesive into neutral HNDs with subsequent ionization. The most abundant peaks are phosphorus clusters but also peaks of impurities from residual gas (e.g. $N_2$ and $H_2O$) and complexes of P with O are visible, most likely originating from phosphates and phosphorus pentaoxide of Resbond® 920. The spectrum shows a higher yield of even-numbered clusters compared to odd-numbered clusters. Two exceptions are $P_5^+$ and $P_7^+$ that show comparable of even higher yields than their respective neighbors $P_6^+$ and $P_8^+$. For clusters sizes n ≥ 26, a drop in ion yield after $P_{4k+2}$ steps ($P_{26}^+$, $P_{34}^+$, $P_{42}^+$) is visible. Furthermore, the ion yield of $P_{24}^+$ exceeds that of $P_{22}^+$, suggesting notable stability for that species.

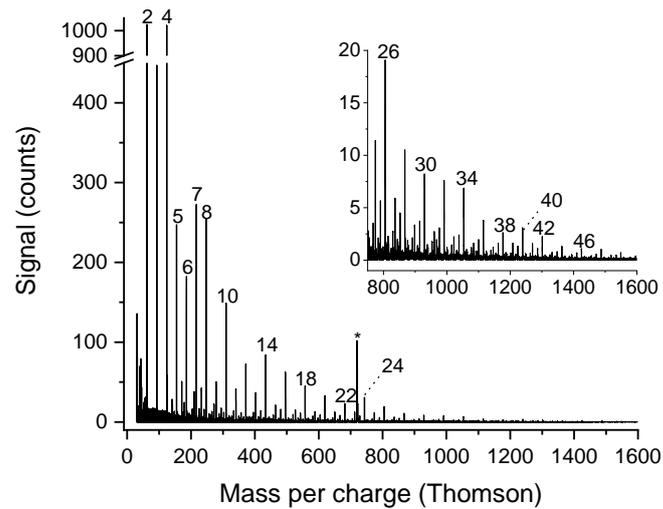

Figure 2: Mass spectrum of cations formed upon electron ionization of HNDs doped with phosphorus vapor emitted from RESBOND® 920 Alumina Ceramic Adhesive. HNDs were produced at a He Pressure of 2.4 MPa and a nozzle temperature of 9.99 K (ionization energy 80 eV). The yield of even-numbered cluster ions is larger than that of neighbouring odd-numbered ones with the exception of $P_5$ and $P_7$. *Peak of residual $C_{60}$. Experimental setup "ClusTOF".

Figure 3 shows the spectrum obtained from cluster formation upon doping P into positively charged, size selected HNDs. The intensity of the even-numbered clusters exceeds the intensity of odd-numbered clusters even more than in the previous spectrum. However, for both the even-numbered series and the odd-numbered series a pattern with intensity anomalies for the addition of $P_4$ units are visible. For odd-numbered clusters, $P_{4k+1}^+$ are particularly abundant. The even numbered clusters exhibit particularly intense clusters for $P_{4k+2}^+$ species up to $P_{18}^+$. However, from $P_{24}^+$ onwards, ions of the composition $P_{4n}^+$ are exceedingly abundant.

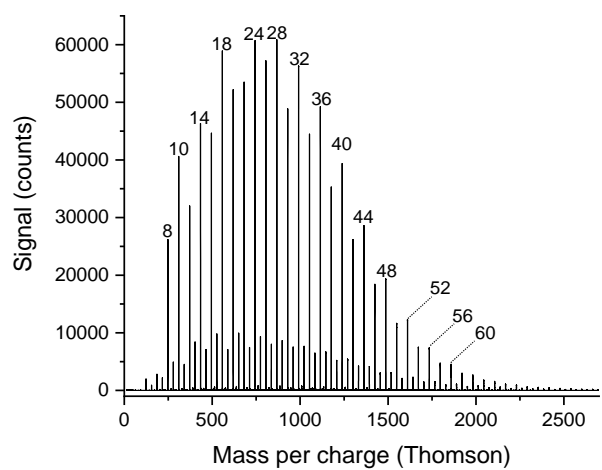

Figure 3: Mass spectrum of cations formed upon doping the vapor of heated (690 K) red phosphorus into charged HNDs that were size selected for approximately 2.5×10$^5$ He/z. HNDs were produced at a He Pressure of 2.0 MPa and a nozzle temperature of 9.2 K (ionization energy 80 eV). Even-numbered clusters are predominant, but odd-numbered $P_{2k+1}^+$ cluster ions remain still clearly visible. Experimental setup "Toffy".

Figure 4 shows the spectrum obtained from cluster formation upon doping P into positively charged, size selected HNDs that were pre-doped with $H_2$. Clusters of the composition $P_{4m+2}H^+$ are particularly abundant, especially up to $P_{42}H^+$. Clusters with an odd number of P atoms are extremely scarce.

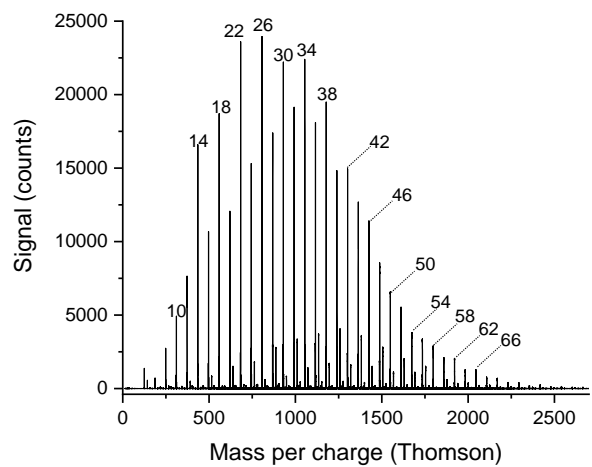

Figure 4: Mass spectrum of cations formed upon pickup of vaporized red phosphorus (690 K) into charged HNDs pre-doped with hydrogen. Proton transfer from $H_{2n+1}^+$ to the first phosphorus dopant (most likely $P_4$) forms protonated phosphorus species $P_mH^+$. HNDs that were size selected for approximately $2.5 \times 10^5$ He/z. HNDs were produced at a He Pressure of 2.0 MPa and a nozzle temperature of 9.2 K (ionization energy 80 eV). Even-numbered clusters are completely dominant, odd-numbered clusters are hardly visible. Experimental setup "Toffy".

Figure 5 and 6 show a comparison of the data of this work (ClusTOF, Toffy, Toffy H) with data from the literature. For these graphs, the normalized signals (0-100) were plotted logarihmically and shifted by a constant value to each subsequent spectrum. Figure 5 shows the comparison for small cluster (12 ≤ n ≤ 24). All phosphorus cluster ions formed in HNDs show a higher abundance of even-numbered clusters, whereas data from Kong and Martin show a higher abundance of odd-numbered clusters. Bulgakov et al. used two different methods of cluster ion formation. When ions were formed directly upon laser ablation, odd-numbered clusters are more abundant (labelled Bulg.00 in the graph [11]). In contrast, when neutral cluster formed upon laser ablation are subsequently ionized by elecron impact ionization, even-numbered clusters were predominant (labelled Bulg.02 in the graph [13]). Figure 6 shows the comparison for larger clusters (25 ≤ n ≤ 67). Also here, cluster ions formed in HNDs show higher abundance of even-numberd clusters in contrast to cluster ions formed by other means. Especially for the data from Bulgakov (Bulg.00) and Huang, cluster ions of the species $P_{8k+1}^+$ are noteably prominent.

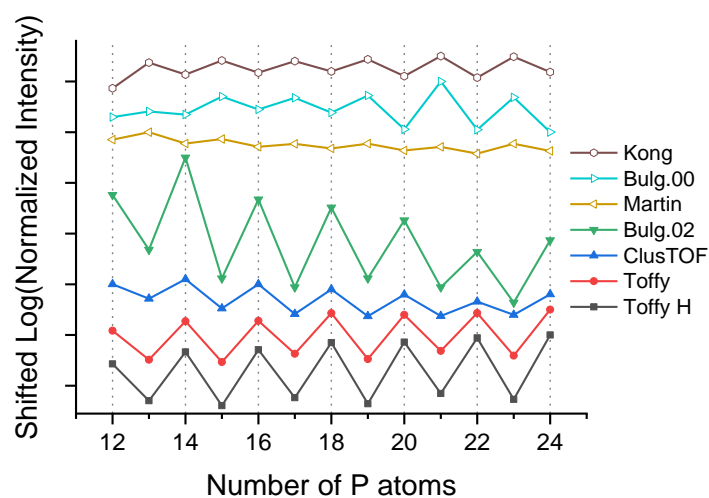

Figure 5: Comparison of P cluster distributions obtained from various measurements for cluster sizes from 12 to 24. Measurements with soft ionization methods (proton transfer – "Toffy H", charge transfer from He (Toffy and ClusTOF)) and data from Bulgakov02 [13] show odd-even oscillations with even-numbered clusters being more intense. Data from Martin [6], Bulgakov00 [11], and Kong [14] show odd-even oscillations with odd numbered clusters being more intense.

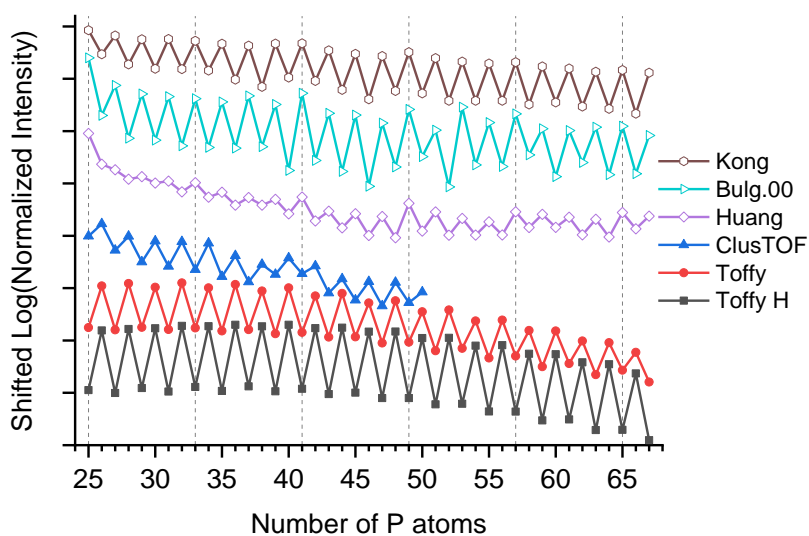

Figure 6: Comparison of P cluster distributions obtained from various measurements for cluster sizes from 25 to 67. Measurements with soft ionization methods (proton transfer – "Toffy H", charge transfer from He (Toffy and ClusTOF)) show odd-even oscillations with even-numbered clusters being more intense. Data from Martin [6], Bulgakov00 [11], Huang [7], and Kong [14] show odd-even oscillations with odd numbered clusters being more intense. These measurements also show magic numbers every 8 P atoms for $P_{33}$, $P_{41}$, $P_{49}$, $P_{57}$, and $P_{65}$.

## Discussion

Depending on the means of cluster formation and ionization, P cluster series exhibit vastly different traits. Vaporization of P occurs mainly in even-numbered $P_2$ and $P_4$ units, resulting in even-numbered aggregates. However, for cationic P odd-numbered closed-shell ions are electronically more stable. It stands to reason that odd-numbered clusters form as more stable products upon the decay of previously formed, even-numbered clusters.

Most results in the literature report a higher abundance for these closed-shell ions after electron ionization of neutral phosphorus clusters formed via gas aggregation [6], ionization upon laser ablation [11], and MALDI TOF analyses [14]. By temporal separation of laser ablation and ionization, Bulgakov et al. reported higher abundance of even-numbered clusters [13]. They hypothesized the formation of particularly stable structures such as even-numbered polyhedra.

The low temperature of HNDs and their high cooling power [44] are able to efficiently quench excess energy from various processes leading to the formation of cluster ions inside. Inelastic collisions of dopant vapor with HNDs and the binding energy of cluster formation introduce energy into the HNDs that leads to vibrational excitation and is released upon evaporation of He atmos. Each eV of internal energy results in the evaporation of about 1600 He atoms, which is determined by the binding engery of a He atom to a HND, i.e., 0.616 meV [45]. Due to the enormous size of HNDs containing millions of He atoms compared to the dopants, the probability for an electron to ionize the dopant directly is negligibly small. The dominant ionization mechanism for heliophilic dopants[28] is charge transfer from an initially formed $He^+$ or a small $He_n^+$ cluster. The difference in the ionization energies of $P_m$ and $He_n$ (n≥1) is in the order of 15 eV, which is completely transferred into the dopant cluster ion. All size distributions of charged dopant cluster $X_n^+$ formed upon electron ionization of neutral doped HNDs exhibit the same intensity anomalies as $X_n^+$ formed by conventional techniques [46]. Therefore, we conclude that some of the excess energy released by the ionization process leads to the evaporation of dopant monomers X. Possible sources for this excess energy are the difference in the ionization energies mentioned above and the enhanced binding energy of X to a charged dopant cluster (solvation energy). In the case of molecular dopants, besides the monomer evaporation also fragmentation of monomers has been reported, although often slightly suppressed compared to electron or photoionization of isolated molecules [47-49]. For molecular decay, reactions that require substantial rearrangement of the molecular constituents can be completely quenched. This includes even highly exothermic reactions, such as the decomposition of trinitrotoluene upon low-energy electron attachment [50]. The vapor of red phosphorus at the presently used temperatures consists predominantly of $P_4$ and a small amount of $P_2$ units. Odd numberd $P_3$ and atomic phosphorus can be neglected. However, electron ionization at 70 eV electron energy of neutral $P_4$ leads to roughly 5% $P_3^+$ and 7% $P^+$ formation.

In the experimental setup, where phosphorus vapor is introduced into pristine charged HNDs (Toffy), the collision of the first $P_4$ or $P_2$ species with a charge center $He_n^+$ is the only process that can lead to a molecular breakup. Pickup and attachment of further $P_2$ or $P_4$ units to these dopant ions will not change the ratio of odd to even-numbered $P_m^+$ cluster ions. Furthermore, this ratio should be constant for all neighboring peaks in the mass spectra. The experimentally determined odd/even ratio is 0.16 (see Table 1). This indicates that charge transfer from $He_n^+$ to $P_4$ is more violent than electron ionization which leads to only 12% odd-numbered fragments[5]. A constant odd-even ratio as a function of the cluster size is indicated by a constant displacement of neighboring peak intensities in a plot with a logarithmic y-axis. According to Figs. 4 and 5, the odd/even ratio stays almost constant up to a cluster size of n=50 and then drops slightly for larger cluster sizes (red solid circles). In contrast, the protonated phosphorus cluster ions exhibit a constant odd-even ratio of 0.01 throughout the complete mass range (black solid squares, Figs. 4 and 5). Predoping charged HNDs with hydrogen (Toffy H) results in proton transfer from an odd-numbered hydrogen cluster cation to the first phosphorus dopant and any excess energy due to the differences in the proton affinities of the reaction partners will be quenched by the surrounding He matrix, as demonstrated recently for various dopants [41]. Thus, this ionization process will not produce odd-numbered fragments and so aggregation of additional even $P_2$ and $P_4$ units explains the extreme dominance of even numbered cluster ions.

Finally, ionization of neutral phosphorus clusters formed upon pickup into neutral HNDs upon charge transfer from $He_m^+$ (ClusTOF) leads to cationic $P_n^+$ clusters with an odd/even ratio of 0.43. In the case of pickup of phosphorus vapor into pristine charged HNDs (Toffy), the very same ionization mechanism is operative and so the three times higher odd/even ratio requires an additional explanation. Laimer et al. demonstrated that multiply charged HNDs are stable once they exceed a critical size and that ionization processes lead to a negligible mass loss[39]. In other words, the ions typically observed via mass spectrometry upon electron ionization of HNDs have to be ejected via Coulomb repulsion from neighboring charge centers or some other exothermic process. This reduces the interaction with the cold He matrix and in the case of dopant cluster ions it can lead to the evaporation of monomers or other weakly bound adducts. This agrees well with the presence of magic numbers in dopant cluster size distributions [46] obtained via this ionization method. Separation of fragments due to molecular dissociation can be suppressed in a cluster environment by so-called caging [51, 52]. In the mass spectra, these trapped fragments cannot be distinguished from stoichiometric parent ions. In ClusTOF, where low-mass ions are quickly ejected from the droplet, such loosely bound complexes are likely destroyed. However, in Toffy, stabilized fragments have a good chance to survive. This readily explains the higher abundance of even-numbered phosphorus cluster ions in Toffy. Gentle removal of the surrounding He matrix via collisions with room temperature He gas makes these cluster ions accessible to mass spectrometric analysis [42].

The odd-even ratio obtained for ClusTOF (electron ionization of neutral HNDs doped with phosphorus) is still lower than for electron ionization of neutral pristine phosphorus clusters formed upon laser ablation (0.71) [13] and much lower than for electron ionization of neutral $P_n$ clusters formed upon gas aggregation of vapor sublimated upon heating of red phosphorus (2.39) [6]. This indicates reduced fragmentation, either by stabilizing $P_4^+$ or caging of odd numberd fragments into a weakly bound $P^+$-$P_3$ complex. Mass spectra of cations formed directly upon laser ablation of red phosphorus exibit the larges odd/even ratios up to 3.87 [14].

Table 1: Ratio of intensity of odd-numbered P cluster ions to even-numbered cluster ions from different measurements.

| Measurment | Odd/Even Ratio |
| --- | --- |
| Toffy H | **0.01** |
| Toffy | **0.16** |
| ClusTOF | **0.43** |
| Bulgakov02 [13] | **0.71** |
| Huang [7] | **0.97** |
| Martin [6] | **2.39** |
| Bulgakov00 [11] | **2.76** |
| Kong [14] | **3.87** |

## Conclusions

We demonstrated that the size distribution of cationic phosphorus clusters strongly depend on the method of production. Whereas most literature data show odd-even oscillation with a high prevalence of closed shell, odd-numbered cations $P_{2k+1}^+$, our data show distinct dominance of even-numbered cations $P_{2k}^+$. Decreasing the input of energy into the formed cluster ions by softer methods of ionization leads to a further reduction of the abundance of odd-numbered clusters. This leads us to the conclusion that cluster formation in HNDs with soft ionization via proton transfer emulates the distribution of neutral phosphorus clusters formed by adding even-numbered $P_{2n}$ units. Harsher cluster formation and ionization conditions on the other hand lead to a size distribution determined by the greater thermodynamic stability of closed-shell, odd-numbered cationic phosphorus clusters.

## Acknowledgements

This work was supported by EFRE (K-Regio project FAENOMENAL, Grant No. EFRE 2016-4), the Austrian Science Fund FWF (Project No. P31149, I4130, P26635), and KKKÖ (Commission for the Coordination of Fusion Research in Austria at the Austrian Academy of Sciences (ÖAW) Grant No. MG 2018-3.